\documentclass[a4,letterpaper,11pt]{article}

\pdfoutput=1 

\usepackage{jheppub} 
\usepackage[T1]{fontenc}
\usepackage{lmodern}
\usepackage{amsmath}
\usepackage{amssymb}
\usepackage{graphicx}
\usepackage{xspace}
\usepackage{slashed}
\usepackage{multirow}

\usepackage{epstopdf}

\newcommand{\nn}{\nonumber}

\newcommand{\be}{\begin{equation}}
\newcommand{\ee}{\end{equation}}
\newcommand{\bea}{\begin{eqnarray}}
\newcommand{\eea}{\end{eqnarray}}

\newcommand{\lp}{\Big{(}}
\newcommand{\rp}{\Big{)}}
\newcommand{\lb}{\Big{\lbrack}}
\newcommand{\rb}{\Big{\rbrack}}
\newcommand{\lbc}{\Big{\lbrace}}
\newcommand{\rbc}{\Big{\rbrace}}

\allowdisplaybreaks[1]

\begin{document}
\preprint{LA-UR-18-26754}

\title{Transverse Momentum Spectra at Threshold for Groomed Heavy Quark Jets}
\author[a]{Yiannis Makris,}
\author[a]{Varun Vaidya}

\affiliation[a]{Theoretical Division, Group T-2, MS B283, Los Alamos National Laboratory, \\
  P.O. Box 1663, Los Alamos, NM 87545, U.S.A. }

\emailAdd{yiannis@lanl.gov}
\emailAdd{vvaidya@lanl.gov}

\abstract{We present the transverse momentum spectrum for a heavy hadron at threshold in a groomed jet initiated by a heavy quark. The cross section is doubly differential in the energy fraction of an identified heavy hadron in the jet and its transverse momentum measured with respect to the groomed (recoil free) jet axis. The grooming is implemented using a soft-drop grooming algorithm and helps us in mitigating the effects of Non-Global logarithms and pile up. For the particular case of a $B$ meson, we identify two distinct regimes of the transverse momentum spectrum and develop an EFT within the formalisms of Soft Collinear Effective Theory (SCET) and Heavy Quark Effective Theory (HQET) for each of these regions. We show how each region can be matched smoothly into the other to provide a prediction for the perturbative transverse momentum spectrum. The EFT also predicts the scaling behavior of the leading non-perturbative power corrections and implements a simple shape function to account for hadronization. We work in the threshold region where the heavy hadron carries most of the energy of the jet since in this regime, we have a very good discriminating power between heavy quark and gluon initiated jets. We observe that the shape of the spectrum is independent of the energy of the jet over a large range of transverse momentum. We propose that this spectrum can be used as a probe of evolution for heavy quark TMD fragmentation function. At the same time, it can be treated as a jet substructure observable for probing Quark-Gluon Plasma (QGP). }

\maketitle

\section{Introduction}
The transverse momentum spectrum with respect to a chosen jet axis of an energetic or massive color-singlet state has been widely studied in literature and is frequently used in probing quantum chromodynamics (QCD) as well as the factorization theorems that separate the long distance physics from ultraviolet hard processes. Hadron-hadron or hadron-lepton collisions (that is, Drell-Yan like spectra and semi-inclusive production of a hadron in deep inelastic scattering) are often used as a tool to study the three dimensional structure of partons inside a hadron. The infrared physics is usually encoded into distribution functions sensitive to the produced transverse momentum and the energy of the hard color singlet state \cite{Collins:2011zzd,GarciaEchevarria:2011rb,Chiu:2012ir} with appropriate soft factors and subtractions \cite{Manohar:2006nz}. This extends the traditional factorization of hadronic structure in terms of collinear Parton Distribution Functions (PDFs) \cite{Georgi:1977mg,Ellis:1978ty,Collins:1981uw,Collins:1989gx} to what are referred to as Transverse Momentum Dependent PDF's or TMDPDF's. When the color-singlet state is within a final state jet not aligned with the beam (for example, observing the transverse momentum spectrum of a final state hadron within an $e^+e^-$-collision using hemispherical jets \cite{Collins:1981uk}), we can likewise extend the notion of collinear Fragmentation Functions (FFs) to include the relative motion of the hadron with respect to all of the other jet constituents in the form of TMDFF's.

Such fragmentation processes are often studied within a \emph{jet} identified with specific jet algorithms such as anti-kT.\footnote{See Refs.~\cite{,Bain:2016clc, Procura:2009vm,Jain:2011xz,Jain:2012uq,Bauer:2013bza,Dasgupta:2014yra,Baumgart:2014upa,Kaufmann:2015hma,Kang:2016ehg,Kang:2016mcy,Chien:2015ctp,Bain:2016rrv,Dai:2016hzf,Dai:2017dpc,Kang:2017glf,Kang:2017btw,Kang:2017mda} for recent work on fragmentation processes both generating and within jets.} Since jets are simply a pattern of radiation that is most likely to occur, there is some arbitrariness in the definition of the jet that can be used. While all reasonable jet definitions usually group together the same energetic radiation into the jet region, the soft radiation that is included inside the jet varies between the different jet algorithms used and this can spoil the equivalence of the TMD-evolution between final state (TMD-FF's) and initial state processes (TMD-PDF's). 

There are issues related to soft correlations that span the whole event, entangling the pattern of soft radiation within the jet to either the underlying event with multiple parton interactions within the colliding hadrons~\cite{Forshaw:2012bi,Catani:2011st,Rogers:2010dm,Zeng:2015iba,Gaunt:2014ska,Rothstein:2016bsq}, or non-global color correlations arising from out-of-jet radiation radiating back into the measured jet, all of which are color connected back to the hard process~\cite{Banfi:2002hw,Dasgupta:2001sh}. Indeed, both effects could potentially spoil the factorization predictions for TMD-spectra found in~\cite{Kang:2017mda,Kang:2017glf}. Thus naively, one suspects that only the TMD-evolution of fragmentation processes within hemisphere jets at an $e^+e^-$ machine could be tied to the TMD-evolution of the $Z$-boson spectrum. 

Developments in jet substructure\footnote{See Ref.~\cite{Larkoski:2017jix} for a comprehensive review.} have shown that the modified mass drop tagging algorithm (mMDT) or soft-drop grooming procedure robustly removes contamination from both underlying event and non-global color-correlations, see Refs.~\cite{Larkoski:2014wba,Dasgupta:2013via,Dasgupta:2013ihk}, and have been applied to study a wide variety of QCD phenomenology within jets \cite{Larkoski:2017iuy,Hoang:2017kmk,Marzani:2017mva,Dasgupta:2016ktv,Frye:2016okc,Frye:2016aiz,Larkoski:2015lea,Dasgupta:2015lxh,Chien:2014nsa}. Exploiting this fact, we will give a concrete proposal as to how one can observe the TMD spectrum of heavy quarks within these groomed jets, where we specify that we study the transverse momentum of a hadron within the jet with respect to the total momentum of the groomed jet, that is, all particles that pass the mMDT or soft-drop procedure.


Jet substructure observables are playing a significant role in various experiments both at high (LHC) and low energy (e.g. RHIC). The focus is on precision standard model measurements  \cite{Chatrchyan:2012sn,CMS:2013cda,Aad:2015cua} as well as searches for new physics \cite{CMS:2011bqa,Fleischmann:2013woa,Pilot:2013bla,TheATLAScollaboration:2013qia}. At the same time, jet substructure measurements are being used as a probe of the Quark-Gluon Plasma (QGP) medium. QCD jets produced from early stage collisions of beam quarks and gluons from two nuclei play a central role in studying the transport properties of QGP. During their propagation through the hot and dense medium, the interaction between the hard jets and the colored medium will lead to parton energy loss (jet quenching) \cite{Bjorken:1982tu,Gyulassy:1990ye,Wang:1991xy}. There have been several experimental signatures of jet energy loss observed at RHIC and the LHC such as modification of reconstructed jets \cite{Aad:2014bxa,Chatrchyan:2012gt,Chatrchyan:2011sx,Aad:2010bu} and jet substructure \cite{Chatrchyan:2012gw,Chatrchyan:2013kwa,Aad:2014wha} as compared to the expectations from proton proton ($pp$) collisions. Continued progress relies on achieving a deeper understanding of the dynamics of jets, allowing for subtle features in a jet to be exploited. This understanding has progressed rapidly in recent years, both due to advances in explicit calculations \cite{Feige:2012vc,Field:2012rw,Dasgupta:2013ihk,Dasgupta:2013via,Larkoski:2014pca,Dasgupta:2015yua}  as well as due to the development of techniques for understanding dominant properties of substructure observables using analytic \cite{Walsh:2011fz,Larkoski:2014gra,Larkoski:2014zma} and machine learning \cite{Cogan:2014oua,deOliveira:2015xxd,Almeida:2015jua,Baldi:2016fql,Guest:2016iqz,Conway:2016caq,Barnard:2016qma} approaches.

While the phenomenology of jets initiated by light/massless partons has been studied extensively (\cite{Feige:2012vc, Field:2012rw, Dasgupta:2013ihk, Dasgupta:2013via, Larkoski:2014pca, Dasgupta:2015yua}, little attention has been paid to the case of jets initiated by massive quarks. For recent work on jet substructure calculations on top quark jets, see \cite{Hoang:2017kmk,Fleming:2007xt} while the issue of heavy quark jet fragmentation has been addressed in \cite{Dai:2018ywt}.
The focus of this paper will be to develop a factorization theorem to study the transverse momentum spectrum of a heavy hadron inside a groomed jet initiated by a heavy quark. We are interested in the regime where the heavy hadron carries most of the energy of the jet.
The introduction of mass as a relevant perturbative scale radically changes the structure of radiation and hence the transverse momentum spectrum of a hadron.
The form of the factorization now involves a TMD Heavy Quark Fragmentation function (written with the formalism of Heavy Quark Effective Theory (HQET)), whose evolution bears little resemblance to the corresponding massless object. Indeed, one of the most startling predictions we make in this paper is that the shape of  transverse momentum spectrum at low (with respect to the hard process) $q_{\perp}$ is essentially independent of the hard scattering scale.

The outline of this paper is as follows. In Section \ref{fact} we develop the factorization theorems in two distinct regimes of transverse momenta within the SCET \cite{Bauer:2000ew, Bauer:2000yr, Bauer:2001ct, Bauer:2001yt, Bauer:2002nz} and HQET \cite{Manohar:2000dt} formalisms. The anomalous dimension (given in Appendix \ref{oneloop}) are then used to resum large logarithms in the transverse momentum in Section \ref{resum}.  The parton and hadron level spectrum that we obtain is compared against \textsc{Pythia} (Section \ref{pythia}). We then provide an analysis of the scaling behavior of the leading non-perturbative effects in Section \ref{np}. We conclude in Section \ref{conc}. The definitions of the operators that appear in our factorization theorem along with the one loop results are provided in Appendix \ref{oneloop}. In Appendix~\ref{sec:factorization} we show some details on the factorization of the cross section.

\section{Factorization}
\label{fact}
Consider the simple case of $e^+ e^-$ collisions where the hard annihilation creates a heavy quark, anti-quark pair. The center of mass energy is much larger than the mass of the quark so that we get two highly boosted partons moving back to back. We isolate a hemisphere jet initiated by heavy quark using a suitable jet algorithm. The heavy quark showers and eventually hadronizes. The jet is groomed with a soft-drop grooming algorithm. The heavy hadron is identified and we measure the energy fraction, $z$, of the hadron with respect to the jet energy. At the same time, we also measure its transverse momentum ($q_{\perp}$) with respect to the axis of the groomed jet (which is recoil free).  We are interested in the regime where the heavy hadron carries most of the jet energy ($z \sim 0.8$). 

\subsection{Soft-Drop Grooming algorithm}
\label{soft-drop}
Soft-drop grooming \cite{Larkoski:2014wba} removes contaminating soft radiation from the jet by constructing an angular ordered tree of the jet, and removing the branches at the widest angles which fail an energy requirement. The angular ordering of the jet is constructed through the Cambridge/Aachen (C/A) clustering algorithm \cite{Ellis:1993tq,Catani:1993hr,Dokshitzer:1997in,Wobisch:1998wt,Wobisch:2000dk}. As soon as a branch is found that passes the test, it is declared the groomed jet, and all the constituents of the branch are the groomed constituents. At the end of the grooming procedure only the narrow energetic core  remains from the original jet. Since at large angles, all collinear energetic radiation is to be found at the center of the jet, no cone is actually imposed to enclose this core. One simply finds the branch whose daughters are sufficiently energetic. Formally the daughters could have any opening angle, though their most likely configuration is collinear.

The strict definition of the algorithm is as follows. Given an ungroomed jet, first we build the clustering history by starting with a list of particles in the jet. At each stage we merge the two particles within the list that are closest in angle\footnote{This merging is usually taken to be summing the momenta of the particles, though one could use winner-take-all schemes \cite{Salam:WTAUnpublished, Bertolini:2013iqa,Larkoski:2014uqa}.}. This gives a pseudo-particle, and we remove the two daughters from the current list of particles, replacing them with the merged pseudo-particle. This is repeated until all particles are merged into a single parent. Then we open the tree back up. At each stage of the declustering, we have two branches available, label them $i$ and $j$. We require:
\begin{align}
  \label{SD:condition}
  \frac{\text{min}\{E_i,E_j\}}{E_i+E_j}>z_{cut}\left(\frac{\theta_{ij}}{R}\right)^{\beta},
\end{align}
where $z_{cut}$ is the modified mass drop parameter, $\beta$ is the parameter which controls the angularities, $\theta_{ij}$ is the angle between $i^{th}$ and $j^{th}$ particle, $R$ is the jet radius and $E_i$ is the energy of the branch $i$. If the two branches fail this requirement, the softer branch is removed from the jet, and we decluster the harder branch, once again testing Eq.~(\ref{SD:condition}) within the hard branch. The pruning continues until we have a branch that when declustered passes the condition Eq.~(\ref{SD:condition}). All particles contained within this branch whose daughters are sufficiently energetic constitute the groomed jet. Intuitively we have identified the first genuine collinear splitting.

For a hadron-hadron collision, one uses the transverse momentum $(p_T)$ with respect to the beam for the condition of Eq.~(\ref{SD:condition}),
\begin{align}\label{SD:condition_pp}
  \frac{\text{min}\{p_{Ti},p_{Tj}\}}{p_{Ti}+p_{Tj}}>z_{cut}\left(\frac{\theta_{ij}}{R}\right)^{\beta}.
\end{align}

We formally adopt the power counting $z_{cut} \ll 1$, though typically one chooses $z_{cut} \sim 0.1$. See \cite{Marzani:2017mva} for a study on the magnitude of the power corrections with respect to $z_{cut}$ for jet mass distributions. Also for the example we considering here we take $\beta = 0$.
\subsection{Momentum modes}

A constrain on the collinearity of the relevant modes is imposed by the fact that we wish to avoid the effects of non-global logs. These will appear if any of our modes contributing to the measurement become sensitive to the boundary of the jet. Since we assume large jet radius of order 1, the only modes that can be sensitive to the boundary are wide angle modes, i.e., modes for which $\theta \sim 1$.Thus to nullify the effect of non-global logs, we need $q_{\perp}/(E_J(1-z))  \ll 1$, where $E_J$ is the energy of the jet, so that all our modes(contributing to the measurement) are collinear and cannot resolve  the boundary.

In our current hierarchy, any large angle ($\theta \sim 1$) mode which hopes to pass soft-drop would scale (in light cone co-ordinates) as $ E_J z_{cut} (1, 1, 1)$. However since this mode have parametrically large transverse momentum, it must necessarily fail soft-drop. So any correction to the cross section from this region of phase space, will only be a $z_{cut}$(we will assume $\beta=0$ for this paper) dependent normalizing factor, but will not influence the shape of the transverse momentum distribution. We left the details of factorization in Appendix~\ref{sec:factorization} and in this section we only show the final result. The cross section can be factorized in the following way:
\begin{equation}
  \label{eq:masterfactorization}
  \frac{d \sigma}{d^2 \vec{q}_{\perp} dz} = \sigma_0(E_J, z_{cut}) S_G( E_J z_{cut}) \times J ( q_{\perp} ,(1-z), E_J, z_{cut}, m) 
\end{equation}
where $S_{G}$ is the global soft function defined and  evaluated  at NLO in \ref{GSoft}, and $\sigma_0(E_J,z_{cut})$ is an overall factor describing the hard process and contribution from non-global emissions. The jet function, $J$, describes the TMD fragmentation within the hemisphere and is sensitive to  the collinear radiation within the hemisphere and contains all the transverse momentum and energy fraction dependence.  

All the radiation modes that contribute to the measurement must necessarily pass soft-drop. Hence they must have atleast $z_{cut}E_J$ amount of energy. At the same time, all the modes have an upper bound on the energy scaling $\sim E_J (1-z)$ set by the energy carried by the heavy quark. Since we are working in the regime $ z_{cut} \sim 1-z$, this uniquely fixes the energy scaling of $all$ modes to be $ z_{cut}E_J  \sim (1-z) E_J$.

We have one more measurement which is the transverse momentum $q_{\perp}$ with respect to the groomed jet axis. For a mode to contribute to this measurement, we require that the angle it makes with the jet axis scale as $\theta \sim q_{\perp}/((1-z)E_J)$. At the same time, Heavy Quark Effective theory (HQET) tells us that at leading power in the 1/m expansion (m being the mass of the heavy quark), there is a limit on the angle  of any radiation mode with respect to the heavy quark given by $\theta_{min} = m/E_J$. Any radiation off the heavy quark at parametrically lower angles is highly suppressed. We can now consider two cases depending on how $\theta$ compares with $\theta_{min}$.
\begin{itemize}
\item{Region 1,  $\theta \sim \theta_{min}$}:
  This directly tells us that $q_{\perp} \sim m(1-z)$ so that 
  transverse momentum $q_{\perp}$ of the heavy hadron is much smaller than its mass. At the partonic level, we then see that $q_{\perp} \ll m$. 
  We then have a single radiation mode that scales as $ p^{\mu} \sim (Q(1-z), q_{\perp}^2/(Q(1-z)), q_{\perp})$ which contributes to both measurements. Equivalently, we can write the scaling of this mode as $p^{\mu} \sim m(1-z)\left( Q/m, m/Q, 1 \right)$, which can be identified as the boosted soft mode of HQET \cite{Fleming:2007xt} (referred to as the ultra-collinear or u-c mode). For convenience we have defined $Q= 2E_J$.  The jet function is then matched onto boosted HQET and further factorizes as follows:
  \begin{equation}
    \label{eq:R1factorization}
    J ( q_{\perp} ,(1-z), E_J, z_{cut}, m) =  H(m) \times  B^{(\perp)}_+(q_{\perp}, E_J(1-z), E_Jz_{cut}, m z_{cut})  
  \end{equation}
  where $B_{+}^{(\perp)}$ is the boosted HQET jet function (\ref{BHQETI}) with additional transverse momentum measurement and $H(m)$  is the matching coefficient from massive SCET to HQET (\ref{HQETm}). The details for the matching procedure are illustrated in Appendix~\ref{sec:factorization}.
  \begin{figure}
    \centerline{\includegraphics[width = \textwidth]{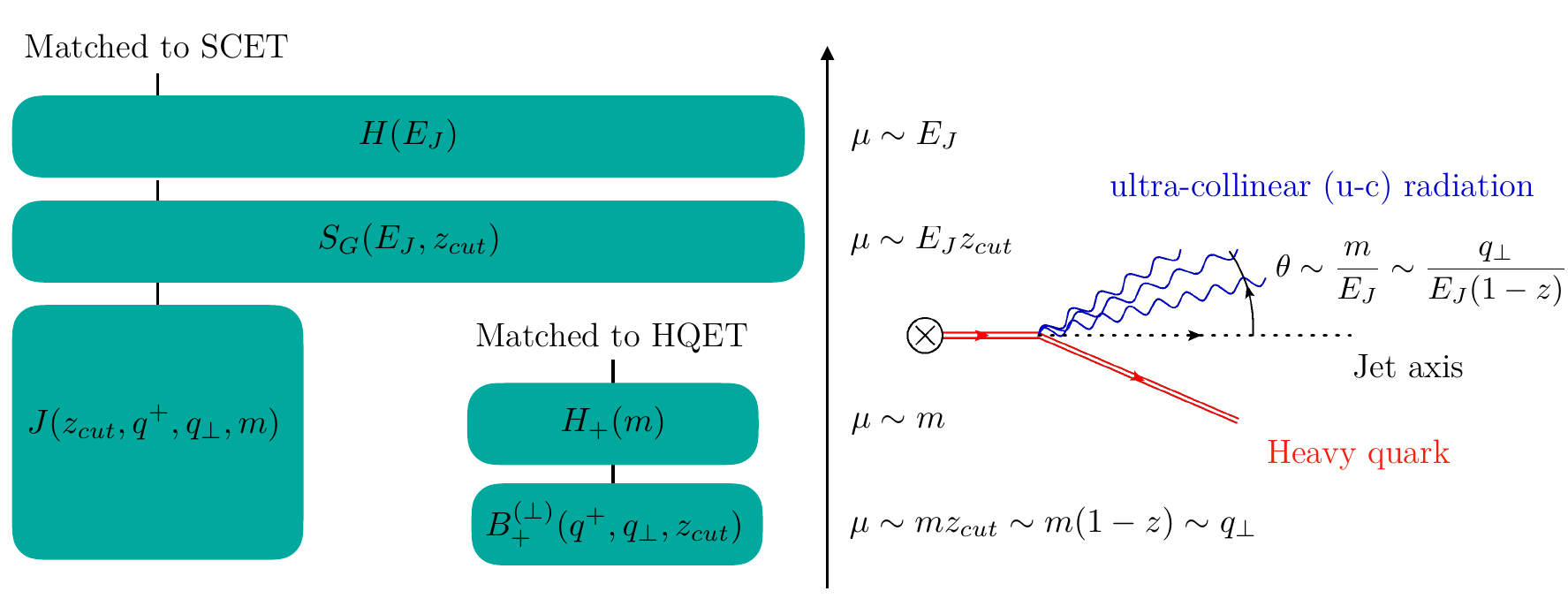}}
    \vskip-0.2cm
    \caption[1]{Hierarchy of scales in region 1 factorization ($q_{\perp} \sim m(1-z)$). There is only a single radiation mode which contributes to both the z and $q_{\perp}$ measurement which can be identified as the boosted soft mode of HQET (referred to as the ultra-collinear or u-c mode)}
    \label{qtsmall} 
  \end{figure}
\item{Region 2, $\theta \gg \theta_{\min}$} : 
  In this case we define two radiation modes, one that contributes to the transverse momentum of the hadron, and another, which does not. The collinear-soft scales as $ p^{\mu} \sim (Q(1-z), q_{\perp}^2/(Q(1-z)), q_{\perp})$ and contributes to both measurements. Our condition on $\theta$ implies that $ q_{\perp} \gg m(1-z)$. The other mode which we refer to as the ultra-collinear(u-c), scales as  $p^{\mu} \sim m(1-z)\left( Q/m, m/Q, 1 \right)$ and only contributes to the $z$ measurement.  The transverse momentum of this mode $p_{\perp} \sim m(1-z) $ is parametrically smaller than $q_{\perp}$ and hence does not contribute to the transverse momentum of the heavy hadron. The jet function is then further factorizes as follows:
  \begin{equation}
    \label{eq:R2factorization}
    J ( q_{\perp} ,(1-z), E_J, z_{cut}, m) = H(m) \times S_C(E_Jz_{cut}, q^+,q_{\perp}) \otimes_{z} B_+( q^+, E_Jz_{cut}, m z_{cut}) 
  \end{equation}
  where $B_{+}$ is the boosted HQET jet function (\ref{BHQETII}) and  $S_C$ is the collinear soft function (\ref{CSoft}). (For the refactorization of the bHQET jet function see discussion in Appendix~\ref{sec:factorization}.) The convolution $\otimes_{z}$ is defined as follows:
  \begin{equation}
    f(q^+)\otimes_{z}g(q^+) = \int_0^{\infty} dq^+\;f(q^+) g(2E_J(1-z) -q^{+})
  \end{equation}
  \begin{figure}
    \centerline{\includegraphics[width = \textwidth]{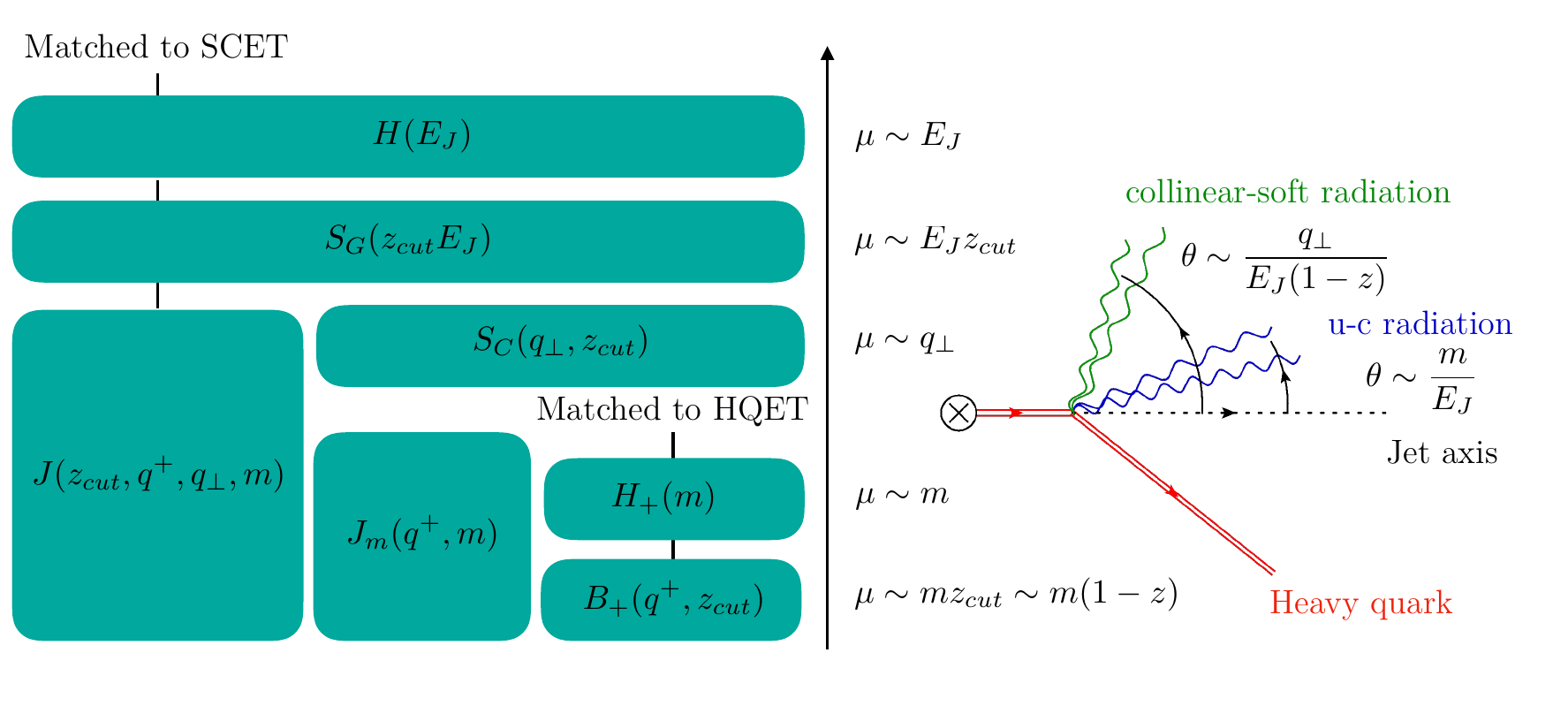}}
    \vskip-0.2cm
    \caption[1]{Hierarchy of scales in region 2 factorization ($q_{\perp} \gg m(1-z)$). We now have an Collinear-soft mode that contributes to both z and $q_{\perp}$ measurements, while the u-c mode only contributes to the z measurement.}
    \label{qtlarge} 
  \end{figure}
\end{itemize}

For a jet energy $\sim$ 100 GeV, we are therefore in the range $q_{\perp} \sim (0, m_b)$. However, the range of $q_{\perp}$ that can be probed by this EFT can be arbitrarily increased as long as the collinearity condition is maintained. Obviously, for a fixed $z$, this can be achieved by increasing the jet energy $E_J$. We demonstrate this in Figure \ref{fig:Q100vs200}.

\section{Resummation}
\label{resum}
We have two distinct regimes of $q_{\perp}$ and hence two separate factorization theorems. We now proceed to resum large logarithms(if any) in each region and match them to obtain the transverse momentum spectrum over the full range of $q_{\perp}$.
\subsection{Region 2: $q_{\perp} \gg m (1-z)$} 
In both regions of phase-space we consider here, the jet function involves the HQET hard coefficient $H_+(m)$. Although this function does satisfy renormalization group equations (RGEs) we will not consider its the evolution since it only contributes to an overall normalization factor and does not influence the shape of the transverse momentum or the energy fraction spectrum.  Therefore here we focus on the RG properties of the collinear-soft function and the HQET jet function which satisfy the following RGE:
\begin{equation}
  \label{eq:measRGE}
  \frac{d}{d\ln \mu } F( E_J(1-z) ) = \int_{0}^{\infty} dq^+ \gamma_{F} (q^+)  F ( 2E_J(1-z)-q^+)
\end{equation}
where $F$ stands for either $S_C$ or $B_+$ and $\gamma_F$ is the corresponding anomalous dimension. We drop the dependence on any  other kinematic variables such as $m,\;z_{cut}$ and $q_{\perp}$ to improve readability. In momentum space the anomalous dimensions read
\begin{align}
  \gamma_{cs} (q^+ ,E_J z_{cut};\mu) & = -2 \frac{\alpha_s(\mu) C_F}{\pi}  \frac{\Theta(q^+ -2E_J z_{cut})}{q^+} \nn \\
  \gamma_{B}  (q^+ ,E_J z_{cut};\mu)   &= \frac{\alpha_s C_F}{\pi} \lbc -2 \ln \lp\frac{\mu}{m z_{cut}} \rp + 1 +2 \frac{\theta( q^+ - 2E_Jz_{cut})}{q^+}\rbc\nn \\ 
  & = (\gamma_J - \gamma_+) \delta(q^+) -  \gamma_{cs} (q^+ ,E_J z_{cut};\mu)
\end{align}
where $\gamma_J$ is the groomed jet anomalous dimension (see Eq.(4.16) in Ref.~\cite{Makris:2017arq}) and $\gamma_+$ the anomalous dimension of the HQET matching coefficient, $H_+(m)$ (see Eq.(\ref{eq:gammaplus})). We emphasize that the non trivial part of the anomalous dimension for the HQET jet function is the same and opposite sign as the one of the collinear-soft function. This is required from the consistency of anomalous dimensions: since the cross section is independent of the factorization scale, the sum of anomalous dimensions should vanish and therefore the non trivial $q^+$ terms in $\gamma_{cs}$ should cancel against the corresponding terms in $\gamma_{B}$. This statement is true for all orders in perturbation theory and here we confirm it at one-loop. 

To solve the RG equation, it is easier to work in Laplace space where
\begin{equation}
  \frac{d}{d\ln \mu }  \tilde{F}( s ) = \tilde{ \gamma}_{F} ( s)  \tilde{F} (s)
\end{equation}
where $\tilde{F}$ denotes the Laplace transform quantity with respect to $q^+$ and $s$ is the Laplace conjugate. The Laplace transform of the collinear-soft anomalous dimension can be written in terms of the incomplete gamma function,  
\begin{equation}
  \tilde{ \gamma}_{cs} ( s, E_J z_{cut}) = -2 \frac{\alpha_s(\mu) C_F}{\pi} \Gamma(0,s Q z_{cut})
\end{equation}
where $Q=2E_J$. The solution to the RGE is,
\begin{equation}
  S_C(s, E_J z_{cut}, q_{\perp};\mu) = \mathcal{U}_{cs} (s;\mu_{sc}, \mu)  S_C(s, E_J z_{cut}, q_{\perp};\mu_{sc})
\end{equation}
where the evolution kernel $\mathcal{U}_{cs}$ is given by
\begin{equation}
  \mathcal{U}_{cs} (s;\mu_{cs}, \mu) = \exp \lb - 2\frac{C_F}{\pi}\Gamma[0,sQz_{cut}]\int_{\mu_{cs}}^{\mu }d\ln \mu' \alpha_{s}(\mu')  \rb
\end{equation}
Similarly for the HQET jet function we have
\begin{equation}
  B_+(s, E_J z_{cut};\mu) = \mathcal{U}_{cs} (s;\mu,\mu_{B}) \exp \lb \int_{\mu_B}^{\mu} d\ln\mu' (\gamma_J-\gamma_+)  \rb   B_{+}(s, E_J z_{cut};\mu_{B}) 
\end{equation}
The scales $\mu_{cs}$ and $\mu_{B}$ are chosen to be the canonical scales for which the logarithms in the fixed order expansion are minimized. As discussed in Appendix~\ref{oneloop} those are,
\begin{align}
  \mu_{cs} &= q_{\perp},\; & \mu_{B} & = m z_{cut}  
\end{align}
Substituting the fixed order result from Eq.(\ref{eq:softfinite})\footnote{Note that the HQET jet function at this order only contributes trivial terms proportional to $\delta^{(2)}(\vec{q}_{\perp})$} and keeping only the non-trivial contributions we have for the LL + LO cross section,
\begin{equation}
  \frac{d\sigma^{(2)}}{dz d^2 \vec{q}_{\perp}}  = \mathcal{N}(E_J,z_{cut},m)\times  \mathcal{L}^{-1} \lb \mathcal{U}_{cs} (s;\mu_{cs}, \mu_B) \frac{\alpha_s C_F}{\pi^2}  \frac{1}{q_{\perp}^2}\Gamma(0, sQz_{cut}) \rb 
\end{equation}
where we absorbed all functions independent of $s$ and $q_{\perp}$ into a single overall normalization factor $\mathcal{N}$. Here $\mathcal{L}^{-1}$ is the inverse Laplace transform with respect to $s$. For LL accuracy we may keep the first term in the QCD $\beta$-function and thus for our final result for the differential cross section in region 2 we get the following,
\begin{equation}
  \label{eq:region1ds}
  \frac{d\sigma^{(2)}}{dz d^2  \vec{q}_{\perp}}  =  \mathcal{N}(E_J,z_{cut},m)\times  \mathcal{L}^{-1} \lb  \frac{1}{q_{\perp}^2}\Gamma(0, sQz_{cut})( r(\mu_{cs},\mu_B))^{-\frac{4 C_F}{\beta_0}\Gamma(0,sQz_{cut})} \rb
\end{equation}
where $r(\mu,\mu_0) = \alpha_s(\mu)/\alpha_s(\mu_0)= 1/(1+\alpha_s(\mu_0)\beta_0/(2\pi)\ln(\mu/\mu_0))$. Note that $ 1/s \sim Q(1-z) \sim Q z_{cut}$ which prevents us from expanding out the incomplete $\Gamma$ function in any small parameter. Also, defining $\tilde s = s Q$, we can write the cross section as 

\begin{equation}
  \frac{d\sigma^{(2)}}{dz d^2  \vec{q}_{\perp}}  =   \mathcal{ \tilde N}(E_J,z_{cut},m)\times  \mathcal{L}^{-1} \lb  \frac{1}{q_{\perp}^2}\Gamma(0, \tilde s z_{cut})( r(\mu_{sc},\mu_B))^{-\frac{4 C_F}{\beta_0}\Gamma(0, \tilde s z_{cut})} \rb
\end{equation}
where $ \mathcal{ \tilde N} =  \mathcal{N}/Q$.
The conjugate variable to $\tilde s$ is just $(1-z)$, so that the shape of the $q_{\perp}$ distribution is independent of the hard scale $Q =2 E_J$. To proceed further, we need to implement the inverse Laplace transform numerically.
For future reference, we define the fixed order cross section in this region ( with the resummation turned off) as
\bea
\label{FO2}
\frac{d\sigma^{(2\text{-FO} )}}{dz d^2  \vec{q}_{\perp}} =    \mathcal{ \tilde N}(E_J,z_{cut},m)\frac{ \alpha_s C_F}{\pi^2} \frac{1}{q_{\perp}^2}\frac{\theta((1-z)-z_{cut})}{(1-z)}
\eea

\subsection{Region 1: $q_{\perp} \sim m(1-z) $}
In this region the collinear-soft and HQET jet function merge into a single function. The corresponding anomalous dimension is given by the sum of $\gamma_{sc}$ and $\gamma_B$. The relevant RGE,
\begin{equation}
  \label{eq:unmeasRGE}
  \frac{d}{d\ln \mu}  B^{(\perp)}_+(\mu)  = \gamma_{B}^{(\perp)} \times  B^{(\perp)}_+(\mu)
\end{equation}
where
\begin{equation}
  \gamma_{B}^{(\perp)} = \gamma_J - \gamma_+ =  \frac{\alpha_s C_F}{\pi} \lbc -2 \ln \lp\frac{\mu}{m z_{cut}} \rp + 1\rbc \nn \\
\end{equation}
Since the anomalous dimension does not depend on either of the measured quantities, there are no convolutions involved and thus the solution can be easily written in momentum space, 
\begin{equation}
  B^{(\perp)}_+(\mu)  = \exp \lb \int_{\mu_B}^{\mu} d\ln\mu' (\gamma_J-\gamma_+)  \rb   B^{(\perp)}_+(\mu_B) \;.
\end{equation}
This allow us to write the resummed cross section as follows:
\bea
\frac{d\sigma^{(1)}}{dz d^2  \vec{q}_{\perp}}  = \mathcal{N}(E_J,z_{cut},m)\times \lb \frac{\alpha_s C_F}{\pi^2}\frac{q_{\perp}^2}{q^+} \frac{\theta( q^+- Qz_{cut})}{ ((q^+)^2m^2/Q^2+q_{\perp}^2)^2} \rb
\eea
where the normalization factor that appears here, $\mathcal{N}$, is the same as in Eq.(\ref{eq:region1ds}).
Notice that there are no large logarithms (involving the measurement scales) to be resummed so the shape is given by the fixed order result. We can also observe that the cross-section goes to zero smoothly as $q_{\perp}$ goes to 0. As in large $q_{\perp}$ region, we can express our cross section factoring out the scale Q
\bea
\frac{d\sigma^{(1)}}{dz d^2  \vec{q}_{\perp}}  = \mathcal{\tilde  N}(E_J,z_{cut},m)\times \lb \frac{\alpha_s C_F}{\pi^2}\frac{q_{\perp}^2}{(1-z)} \frac{\theta( (1-z)- z_{cut})}{ ((1-z)^2m^2+q_{\perp}^2)^2} \rb
\eea
So that once again, the shape of the $q_{\perp}$ distribution is independent of the hard scale.

\subsection{The transverse momentum spectrum}

The two regimes of our resummed cross section must smoothly match into the other. It is clear that the only difference between the two regimes is that the $\ln( m z_{cut}/q_{\perp})$ will be resummed in one ($q_{\perp} \gg m(1-z)$) and not in the other.  On the other hand the power corrections in $ q_{\perp}/(m/Qq^+)$ are important in the low $q_{\perp} $ regimes. So in order to obtain a reliable spectrum in both regions while smoothly interpolating between the two regions we need to turn off the resummation of the $\ln( mz_{cut}/q_{\perp})$ as we approach intermediate $q_{\perp} \sim 2.5$ GeV and match the result to the fixed order (which does not involve any $q_{\perp}$ or $(1-z)$ resummation) cross section. The merging between the two regions can be achieved through multiplicative matching
\begin{equation}
  \frac{d\sigma^{(1+2)}}{dz d^2  \vec{q}_{\perp}} = \frac{d\sigma^{(1)}}{dz d^2  \vec{q}_{\perp}} \times  \frac{d\sigma^{(2)}}{dz d^2  \vec{q}_{\perp}} \Big{/}\frac{d\sigma^{(2\text{-FO} )}}{dz d^2  \vec{q}_{\perp}} 
\end{equation}
where $d\sigma^{(2\text{-FO} )}$ (Eq. \ref{FO2}) is the fixed order result in region 2.  When the resummation in $d\sigma^{(2)}$ is turned off at small $q_{\perp}$ then is easy to show the following asymptotic behavior for $d\sigma^{(1+2)}$
\begin{align}
  \frac{d\sigma^{(1+2)}}{dz d^2  \vec{q}_{\perp}} (q_{\perp} \sim m z_{cut}) &\simeq \frac{d\sigma^{(1)}}{dz d^2  \vec{q}_{\perp}} (q_{\perp})\nn \\
  \frac{d\sigma^{(1+2)}}{dz d^2  \vec{q}_{\perp}} (q_{\perp} \gg  m z_{cut}) &\simeq \frac{d\sigma^{(2)}}{dz d^2  \vec{q}_{\perp}} (q_{\perp})
\end{align}

\begin{figure}[t!]
  \centerline{\includegraphics[width = \textwidth]{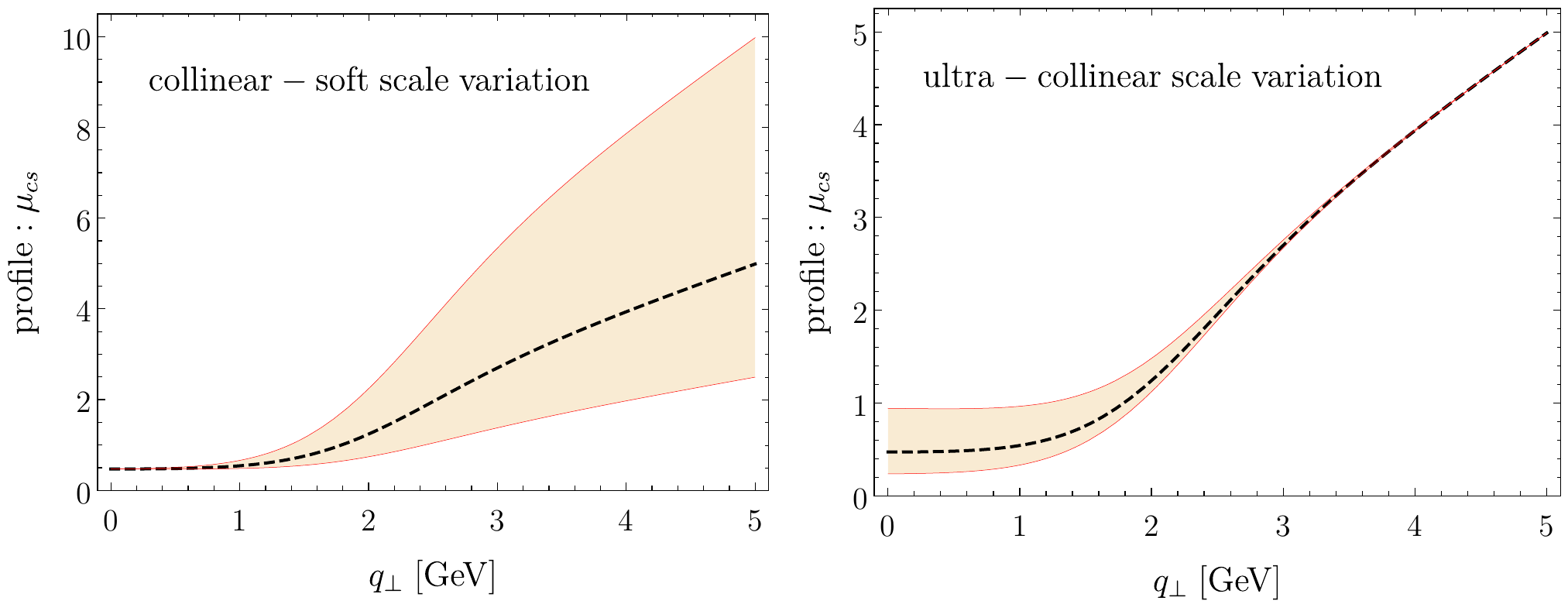}}
  \caption{Profiles in the collinear soft scale $\mu_S$, which smoothly turn off the resummation going from region 2 to region 1. The variation in the profiles probes the error band due to missing higher order corrections}
  \label{fig:profiles}
\end{figure}

Both these regimes have a common resummation factor (resumming double and single logarithms of $Q/mz_{cut}$)  which we can be ignored since it does not affect the shape of the distribution. Turning off of the resummation in region 2 is achieved using profile scales in $\mu_{cs}$ as shown in Figure~\ref{fig:profiles}. At the same time we probe for higher order corrections using scale variations about the central profile by a factor of two and one half.

\section{Comparison with simulation}
\label{pythia}
In this section we compare our LL+LO result against monte-carlo simulations.  For simplicity we consider the  process $e^+ e^- \to b+X$, where we groom with soft-drop the hemisphere which the $b$-quark is found. The two hemispheres are identified by the plane perpendicular to the thrust axis~\cite{Farhi:1977sg}. We measure the  energy fraction, $z$, and the transverse momentum, $q_{\perp}$, of the heavy-quark with respect to the groomed-jet axis\footnote{In this case the groomed-jet axis is defined as the direction of the total three-momenta of all particles in the corresponding hemisphere that pass the grooming procedure.}.  

For the simulations we used Madgraph~\cite{Alwall:2014hca} for generating the LO hard process  $e^+ e^- \to b+\bar{b}$ and then the partonic shower is implemented in \textsc{Pythia} 8~\cite{Sjostrand:2006za,Sjostrand:2007gs}. For the analysis, \textsc{FastJet}~\cite{Cacciari:2011ma} is used along with the corresponding add-ons from \textsc{FastJet}-contrib for imposing soft-drop grooming.

\begin{figure}[t!]
  \centerline{\includegraphics[width = \textwidth]{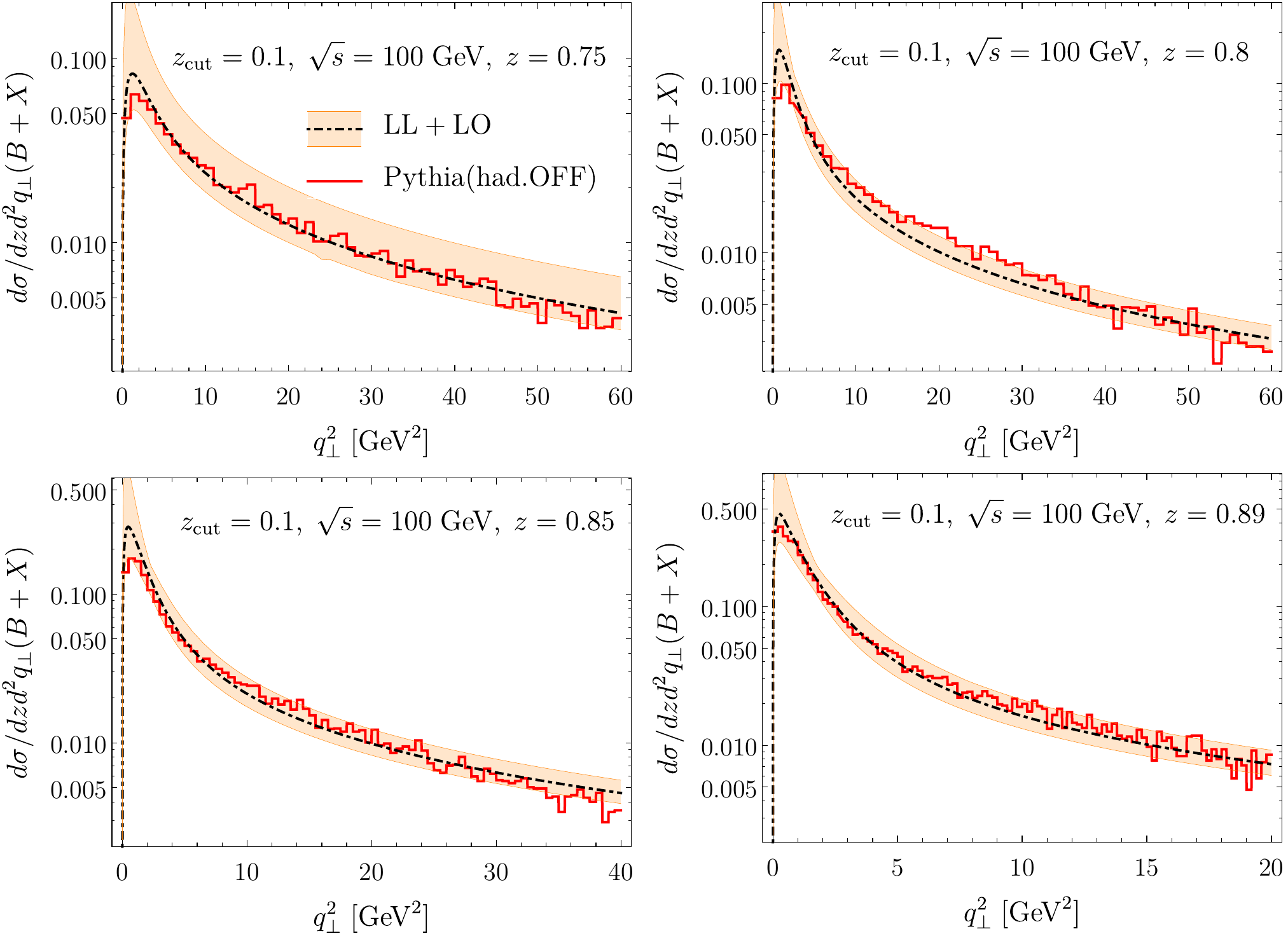}}
  \caption{Comparison of the analytical result(with error bands)  at LL+LO  accuracy with parton level \textsc{Pythia} with a jet energy $E_J =Q/2 \sim 50 GeV$.}
  \label{fig:Q100}
\end{figure}

The kinematic variables we choose for the comparison are for  center of mass energy $\sqrt{s}=100$ GeV  and for the grooming parameter $z_{\text{cut}} =0.1$. In Figure~\ref{fig:Q100} we compare the LL+LO results against the simulation for four different values of the energy fraction: $z=0.75,\;0.80,\;0.85,\;0.89$. We find good agreement with the simulation for most of the range of $q_{\perp}$ and within the theoretical uncertainty. The uncertainty for the analytic result is obtain by varying the jet and soft scales, $\mu_B$ and $\mu_{cs}$ by a factor of 1/2 and 2 around their canonical values. We note that for small transverse momenta,  $q_{\perp}<2$ GeV, the scale variation results in a large uncertainty. This is a result of the small scale $m z_{cut} \sim 0.5$ GeV that dominates the uncertainty band in that region.

\begin{figure}[t!]
  \centerline{\includegraphics[width = \textwidth]{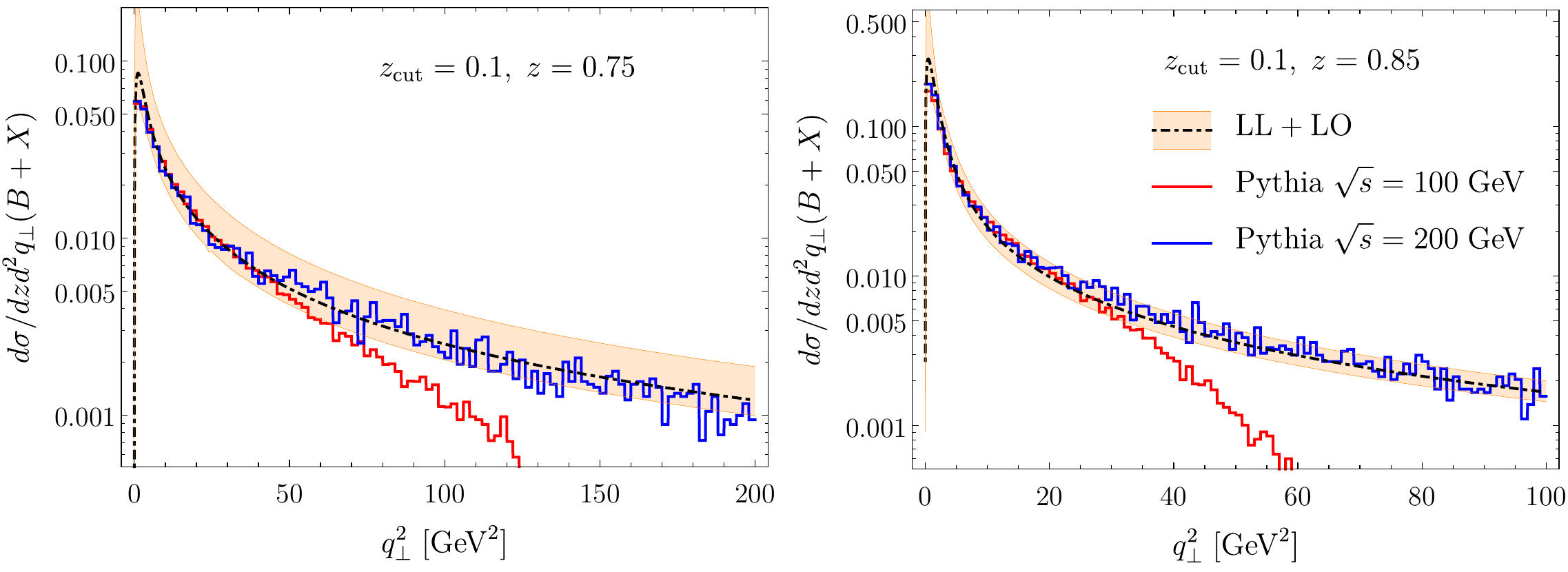}}
  \caption{Comparison with parton level \textsc{Pythia} for different values of jet energy $E_J =Q/2$. As the jet energy increases, the EFT is valid till a higher value of $q_{\perp}$. This plot vividly demonstrates the Q independence of the shape at low $q_{\perp}$. }
  \label{fig:Q100vs200}
\end{figure}

Although according to our result, the shape of the transverse momentum distribution does not explicitly depend on the hard scale $Q$, at larger values of $q_{\perp} \sim Q(1-z)$ power corrections of the form $q_{\perp}/Q(1-z)$  become important. These terms are not captured by the EFTs used here. To test these observation we compare our analytic result against the simulation for $\sqrt{s}=100$ and $200$ GeV. The results are shown in Figure~\ref{fig:Q100vs200} for the cases $z=0.75$ and $z=0.85$. We find that indeed the partonic transverse momentum spectrum is independent of the value of $Q$  and in agreement with our predictions away from the region $q_{\perp} \sim Q(1-z)$.

The simulation results eventually vanish at $q_{\perp}^{max} = E_J (1-z)$. In this region the heavy meson recoils against soft radiation close to the hemisphere boundary. To describe the cross section in this region one can construct an EFT with the following hierarchy:
\begin{equation}
  Q\gg Q(1-z) \sim q_{\perp} \gg m.
\end{equation}
In this region the cross section is mostly insensitive to the grooming procedure and there are two relevant modes that contribute to the measurements. The global soft mode, $p_s^{\mu} \sim Q(1-z)(1,1,1) \sim q_{\perp}(1,1,1)$ which now contribute both to the energy and transverse momentum and the ultra-collinear mode, $p_{uc} \sim m(1-z)(Q/m,m/Q,1)$ which  only contributes to the energy measurement. Note that the naive fixed order EFT result diverge in that region. The correct behavior of the cross section can be reproduced within the EFT only at NLL' (or higher logarithmic accuracy) as described in Ref.~\cite{Larkoski:2014tva}. However, we do not investigate this region further in this paper since we do not go  beyond a one loop calculation. At the same time, this far tail region will be very hard to measure and hence will not be of significant phenomenological importance.

\section{Non-perturbative corrections}
\label{np}
In order to probe the non-perturbative corrections, we lower the virtuality of the modes from $q_{\perp}$ to $\Lambda_{\text{QCD}}$. 
Since the HQET ultra-collinear mode has the lowest virtuality in our scale hierarchy, this function will contribute to the dominant non-perturbative effects. The HQET jet operator in region 1 is defined as Eq.(\ref{b+1})
\bea
B_{+}^{(\perp)} = \langle 0| \bar h_{v_+} W_n \delta( q^+-(1-\hat \Theta_{SD}) \mathcal{P}^-_X)\delta^2(q_{\perp}- (1-\hat \Theta_{SD}) \mathcal{P}_{\perp}) |B+X \rangle \langle B+X |W_n^{\dagger}H_{v_+}|0\rangle \nn\\
\eea
Taking the Laplace and Fourier transform with respect to $q^+$ and $\vec{q}_{\perp}$ we can write 
\bea
B_{+}^{(\perp)} =  \langle 0| \bar h_{v_+} W_n e^{-s (1-\hat \Theta_{SD})p^-_X}e^{\vec{b} \cdot \vec{q}_{\perp}(1-\hat \Theta_{SD})} |B+X \rangle \langle B+X |W_n^{\dagger}h_{v_+}|0\rangle
\eea
We consider the case when $p^-_X, q_{\perp} \sim \Lambda_{\text{QCD}}$, which then induces power corrections of the form $   s \Lambda_{\text{QCD}}, b \Lambda_{\text{QCD}}$. Expanding out and keeping the leading order term, we have
\bea
B_{+}^{(0)} =  \langle 0| \bar h_{v_+} W_n |B+X \rangle \langle B+X |W_n^{\dagger}h_{v_+}|0\rangle
\eea
which is simply a normalization factor. Lets look at the first non trivial power correction. We have two contributions, one of which is from the expansion in $s \Lambda_{\text{QCD}}$.
\bea
P_1= -s p^-_X\langle 0| \bar h_{v_+} W_n  |B+X_{SD} \rangle \langle B+X_{SD} |W_n^{\dagger}h_{v_+}|0\rangle
\eea 
where $X_{SD}$ indicates that this operator exists only for the radiation that passes soft-drop.
To establish some type of scaling universality, we do a Lorentz transformation to the rest frame of the heavy quark. The Wilson line remains unchanged while the velocity of the heavy quark now becomes $v \sim (1,0,0,0)$ This simply gives us 
\bea
\label{power_np}
P_1=-s Q/m  \langle 0| \bar h_{v} W_n  \mathcal{P}^-_X|B+X_{SD} \rangle \langle B+X_{SD} |W_n^{\dagger}h_{v}|0\rangle
\eea
The matrix element $\langle 0| \bar h_{v} W_n  \mathcal{P}^-|B+X_{SD} \rangle \langle B+X_{SD} |W_n^{\dagger}h_{v}|0\rangle$ is independent of the energy of the jet or the mass of the heavy quark and has dimensions of energy. It is however, dependent on the soft-drop condition and hence is sensitive to the value of $z_{cut}$. Typically we would expect this object to have value $\sim \Lambda_{\text{QCD}}$. In order to go deep into the non-perturbative region $ Q(1-z) < \Lambda_{\text{QCD}}$, in principle, we need to have an all orders description of the non-perturbative corrections. The usual way to deal with this is to put in some type of a model shape function that captures the dominant non-perturbative physics. Any such model would also have the constraint that the leading power correction should be of the form Eq.\ref{power_np}. Here we consider a simple exponential model which correctly recovers the leading power correction.
\bea
\tilde{f}_{\text{np}}(s) = e^{ -s Q \Lambda/m }
\eea
Taking the inverse Laplace transform we get 
\bea
f_{\text{np}}(q^+) = \delta\lp q^+ - Q \frac{\Lambda}{m} \rp 
\eea
which depends on a single parameter $\Lambda$ that needs to be fitted from simulation/experiment. The differential spectrum at the hadronic level can then be given as a convolution of this shape function with the partonic jet function.
\begin{equation}
  \label{eq:hadronizationeffects}
  B_{+}^{(\perp)}(E_{J}(1-z)) \Big{\vert} _{\text{had.}} = B_{+}^{(\perp)}(q^+) \otimes_{z} f(q^+) =  B_{+}^{(\perp)}(E_J(1-z- \Lambda/m))
\end{equation}
which implements a simple shift in $z$, so that the hadronized spectrum is same as the partonic one at a lower value of $z$. More sophisticated models exist in literature \cite{Fickinger:2016rfd,Ligeti:2008ac} which smoothen out the delta function about its central value of $y=\Lambda/m$, but have essentially have the same effect.

\begin{figure}[t!]
  \centerline{\includegraphics[width = \textwidth]{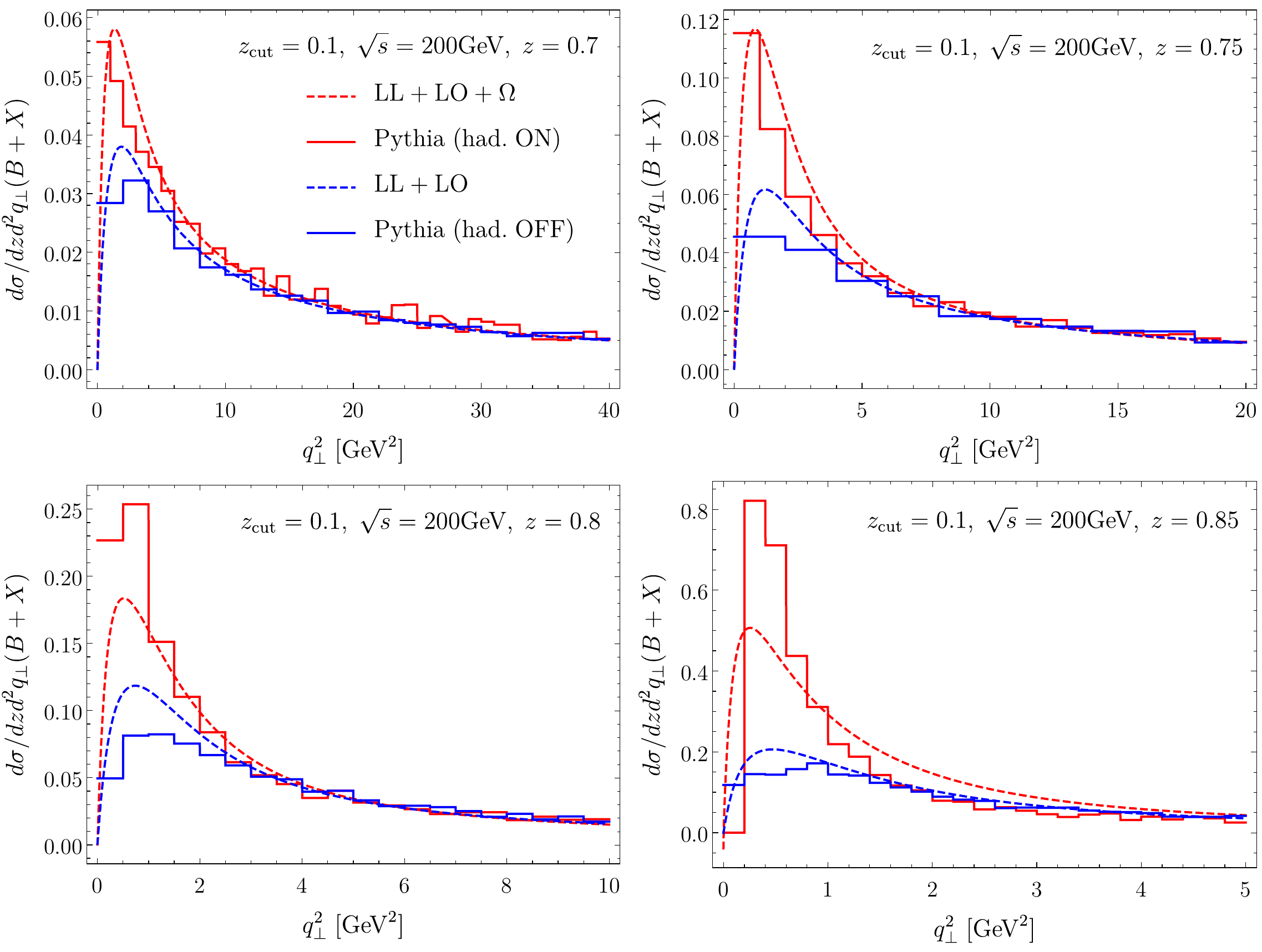}}
  \caption{Hadronization effects in \textsc{Pythia} versus the simplified prescription of Eq.(\ref{eq:hadronizationeffects}). The results are shown for various values of the momentum fraction, particularly we choose $z=0.7,0.75,0.8,$ and $0.85$. We find that, as expected, the simple shift works better for values away from the energy threshold, $1-z_{cut}$. }
  \label{fig:hadON}
\end{figure}

Figure \ref{fig:hadON} shows the comparison of hadronized \textsc{Pythia} with this simple shift model. The blue curves are the parton level distributions, while the red ones are hadron level. We have set the value of $\Lambda \sim$ 0.2 GeV, in order to have a good match with \textsc{Pythia} at lower values of $z$ ($z \sim  0.75$). Note that the parameter $\Lambda$ found here is smaller than the corresponding $\lambda$ found in Ref.\cite{Fickinger:2016rfd}. This is due to the fact that, in our analysis $\Lambda$ depends on soft-drop which suppresses contribution from the wide angle radiation. We see that the agreement with \textsc{Pythia} worsens in at higher $z$ for very low $q_{\perp}$. This is mainly because of the fact that the process of grooming and hadronization do not commute. Looking the $z$ spectrum we see that the number of events at the parton level which just fail soft-drop is very large (This is to be contrasted with a gluon initiate jet where the cross section drops to zero at such high $z$ values). On hadronization, many of these partons acquire enough energy to pass soft-drop, thus increasing the total number of events at low $q_{\perp}$ at the hadronized level. All of these events are, of course, missed by the analytical calculation, which implements hadronization corrections only on those events that pass soft-drop at the partonic level.

We can do a similar boost for the power correction from $q_{\perp}$ which gives us 
\bea
-b^2  \langle 0| \bar h_{v} W_n  \mathcal{P}_{X\perp}^2|B+X \rangle \langle B+X |W_n^{\dagger}h_{v}|0\rangle
\eea
where we have used rotational invariance. In the same vein as the power corrections in z, we can implement an exponential( in this case a Gaussian) model to implement the non-perturbative corrections. We do not explicitly implement these correction in this paper, since the perturbative error bands in the very low $q_{\perp}$ region are large and we are not yet sensitive enough to the non-perturbative effects to extract out a meaningful parameter.
\section{Conclusion}
\label{conc}

We present the transverse momentum spectrum for a heavy hadron identified in a groomed jet. We work in a regime where the heavy hadron carries most of the energy of the jet. The grooming is implemented with a soft-drop grooming procedure. The results presented are for the case of a $b$ quark initiated jet in $e^+ e^-$ collision, although this can be easily extended to the case of $pp$ collisions. The grooming gives us a significant advantage in that we are insensitive to corrections due to non-global logarithms or pile-up.

We work in a hierarchy $E_J \gg E_J(1-z) \sim E_J z_{cut} \gg m \gg \Lambda_{\text{QCD}}$, where $E_J$ is the energy of the jet, $z$, the fraction of the jet energy carried by the heavy hadron, while $z_{cut}$ is the grooming parameter. $m$ is the mass of the heavy quark. We identify two distinct regimes for the transverse momentum $q_{\perp} \sim m (1-z) \ll m $ and $E_J(1-z) \gg q_{\perp} \gg m(1-z) $.

We formulate two separate factorization theorems in these two regimes to capture the dominant contribution to the cross section at leading power in this hierarchy. We resum the large logarithms in $q_{\perp}$ at Leading Log accuracy and smoothly match the cross section in the two regimes of transverse momentum. For the low $q_{\perp}$ regime, the shape of the distribution is given entirely by the fixed order coross section and there are no large logarithms (in $q_{\perp} $ or $(1-z)$) that require resummation.

Comparing with partonic \textsc{Pythia}, we obtain an excellent agreement in the regime where the EFT is valid.  There is deviation in the far tail region $q_{\perp} \sim E_J(1-z)$, where the cross section becomes sensitive to the boundary of the jet. From our analytic calculation, we observe that the shape of the $q_{\perp}$ spectrum is independent of the energy of the jet over most of the $q_{\perp}$ range, again, deviation from this behavior is in the far tail where the angle $ q_{\perp}/(E_J(1-z))$ of the radiation recoiling against the heavy quark nears the edge of the jet.

We obtain analytical expressions for the scaling behavior of the leading non-perturbative power corrections to the cross section. These leading power corrections are incorporated in a model implementing a simple shift in $z$. Comparing the $q_{\perp}$ region, we again obtain good agreement with \textsc{Pythia} except for the very low $q_{\perp}$ region. As explained in Section \ref{np}, this discrepancy, especially at high values of $z$ is due to the large number of events that pass soft-drop after hadronization, which are missing from the analytical calculation.

While our calculation only captures the leading-logarithmic (LL) behavior of the cross section, uncertainty reduction can be obtained by a higher order computation. This will require a two loop computation of the HQET jet functions in each regime of $q_{\perp}$. Especially in the low $q_{\perp}$ region, the error band is large since the fixed order cross section evaluated at the scale $m z_{cut} \sim 0.5$ GeV is essentially dictating the shape of the transverse momentum distribution.

We propose that this observable can be used as a probe of heavy quark TMD fragmentation function. At the same time, we can also use this measurement as a jet substructure observable since it is sensitive to the radiation pattern in the jet.

\begin{acknowledgments}
  We would like to thank Duff Neill and Emanuele Mereghetti for several useful discussions during the course of this project. This work was supported by the U.S. Department of Energy through the Office of Science, Office of Nuclear Physics under Contract DE-AC52-06NA25396 and by an Early Career Research Award, through the LANL/LDRD Program, and within the framework of the TMD Topical Collaboration.

\end{acknowledgments}

\appendix

\section{One loop results}
\label{oneloop}

\subsection{GSoft}
\label{GSoft}
The global soft function is defined as the following matrix element of Wilson lines 
\bea
S_G(z_{cut}) = \frac{1}{N_C} \text{Tr} \langle 0|T\{Y_nY_{\bar n}\}\hat \Theta_{SD}\Theta_R\bar T\{Y_n Y_{\bar n}\}|0\rangle
\eea
$\hat \Theta_{SD}$ denotes the soft drop groomer. $\Theta_R$ imposes the hemisphere jet constraint. We require that the global soft modes fail soft drop. From literature \cite{Frye:2016aiz}, we can write down the result for the one loop singular piece 
\bea
S_G = 1+\frac{\alpha_s C_F}{\pi} \ln^2 \left(\frac{\mu}{Q z_{cut}}\right)
\eea
So the natural scale for this function is $\mu_{gs} = Q z_{cut}$.

\subsection{BHQET}
\label{BHQET}

\subsubsection{Region 1: $q_{\perp} \sim m(1-z)$}
\label{BHQETI}
\begin{figure}[t!]
  \centerline{\includegraphics[width = \textwidth]{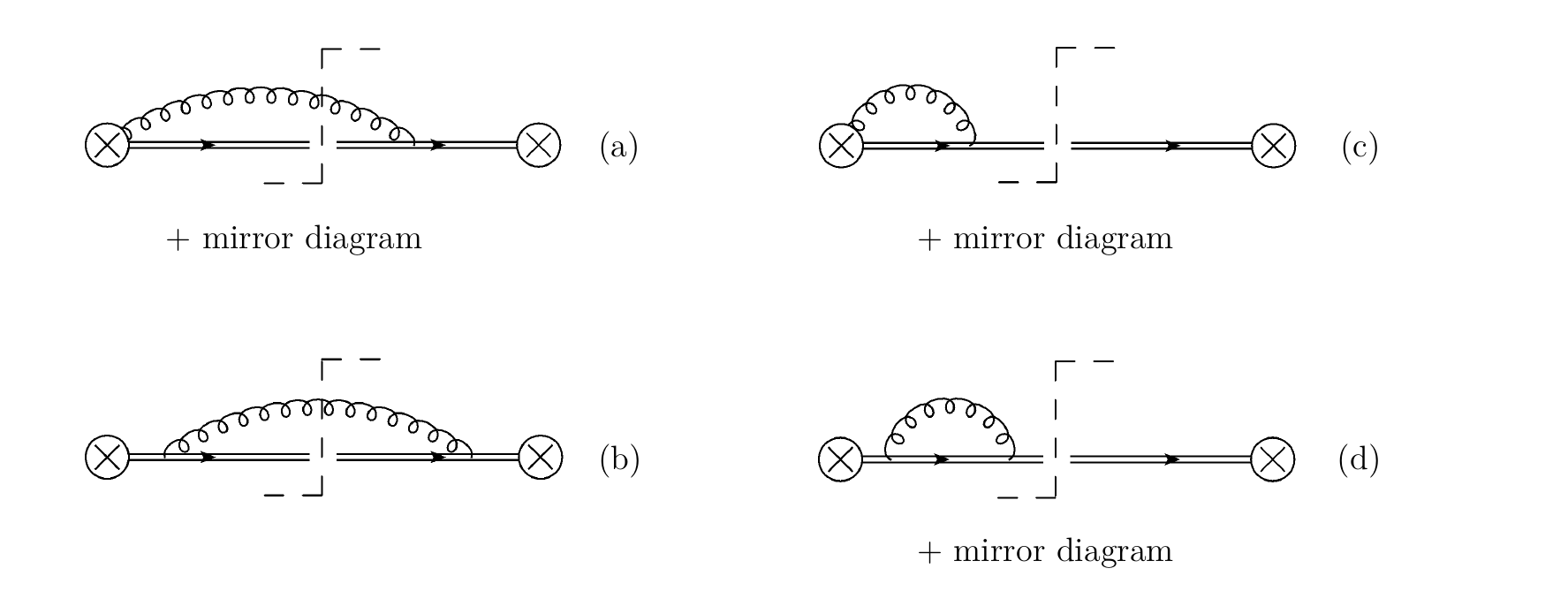}}
  \caption{One loop diagrams for the HQET jet function.}
  \label{fig:diagrams}
\end{figure}

The residual momentum that make up the modes of this region (known as ultra collinear modes) scales as 
\bea
k^{\mu} \sim \Gamma \left(m/Q, Q/m, 1 \right)  
\eea 
where  $\Gamma = m (1-z)$ is the IR scale for this EFT ($m$ being the hard scale). In this regime $q_{\perp} \sim m(1-z)$ so that the jet function contributes to both measurements. Therefore the boosted HQET jet function is defined as 
\bea
\label{b+1}
B_+^{(\perp)} = \frac{1}{\mathcal{N}} \langle 0 | \bar h_{v_+} W_n  \;\mathcal{M}^{SD}(q^+,\vec{q}_{\perp}) \;W_n^{\dagger} h_{v_+} |0 \rangle . 
\eea
where  $v_+ = ( m/Q, Q/m, 0_{\perp})$ is the velocity of the boosted heavy quark and $\mathcal{M}^{SD}$ the measurement  operator
\begin{equation}
  \mathcal{M}^{SD}(q^+,\vec{q}_{\perp}) = \delta \left(q^+-(1-\hat \Theta_{SD})\mathcal{P}^-\right)\delta^2\left(\vec{q}_{\perp}-(1-\hat \Theta_{SD}) \mathcal{\vec P}_{\perp} \right)
\end{equation}
At one loop there are two real emission diagrams (diagrams (a) and (b) in  Figure~\ref{fig:diagrams}).

\bea
B_{+,\;\text{a}}^{(\perp)} &=&  2 g^2 C_F \tilde \mu^{2\epsilon} \int \frac{d^dk}{(2\pi)^{d-1}}\frac{\delta^+(k^2)\delta(\bar n \cdot k- q^+) \delta^2(\vec{q}_{\perp}-\vec{k}_{\perp})}{v_+ \cdot k \bar n \cdot k}\theta( \bar n \cdot k - Qz_{cut})\nn\\
& +& 2 g^2 C_F \tilde \mu^{2\epsilon}\delta(q^+) \delta^2(\vec{q}_{\perp}) \int \frac{d^dk}{(2\pi)^{d-1}}\frac{\delta^+(k^2)}{v_+ \cdot k \bar n \cdot k}\theta( Qz_{cut}-\bar n \cdot k  ) \nn\\
&\equiv& B_{+,\;\text{a-1}}^{(\perp)}+ B_{+,\;\text{a-2}}^{(\perp)}
\eea
\bea
B_{+,\;\text{a-1}}^{(\perp)} =    \frac{2 g^2 C_F}{(2\pi)^3}\frac{1}{q^+} \frac{1}{ (q^+)^2m^2/Q^2+q_{\perp}^2}\theta( q^+- Qz_{cut})
\eea
\bea 
B_{+,\;\text{a-2}}^{(\perp)}  &=&  2 g^2 C_F \tilde \mu^{2\epsilon}\delta(q^+) \delta^2(\vec{q}_{\perp}) \int \frac{d^dk}{(2\pi)^{d-1}}\frac{\delta^+(k^2)}{v_+ \cdot k \bar n \cdot k}\theta( Qz_{cut}-\bar n \cdot k ) \nn\\
&=&  -\frac{g^2 C_F}{8 \pi^2} \left(\frac{\tilde \mu}{m z_{cut}}\right)^{2\epsilon}\delta(q^+) \delta^2(\vec{q}_{\perp})\frac{\Gamma[\epsilon]}{\epsilon}
\eea
Expanding out all the pieces and keeping the singular terms, we get 
\bea
B_{+,\;\text{a}}^{(\perp)} =\frac{\alpha_s C_F}{\pi^2}\frac{1}{q^+} \frac{1}{ (q^+)^2m^2/Q^2+q_{\perp}^2}\theta( q^+- Qz_{cut})-\frac{\alpha_s C_F}{\pi}\ln^2 \left(\frac{\mu}{m z_{cut}}\right)
\eea
which reproduces the correct cusp anomalous dimension. 
We have one more diagram 
\bea
B_{+,\;\text{b}}^{(\perp)} &=& -g^2 C_F \int d^dk \frac{\delta(k^2) (v_+)^2 \delta( q^+-\bar n \cdot k)\delta^2(\vec{q}_{\perp}-\vec{k}_{\perp})}{(v_+ \cdot k )^2}\theta(\bar n \cdot k - Qz_{cut})\nn\\
&-& g^2 C_F \delta( q^+)\delta^2(\vec{q}_{\perp})\int d^dk \frac{\delta(k^2) (v_+)^2 }{(v_+ \cdot k )^2}\theta( Qz_{cut}-\bar n \cdot k)\nn\\
&=&-\frac{2 g^2 C_F}{(2\pi)^3} \frac{m^2}{Q^2}q^+ \left(\frac{1}{m^2/Q^2 (q^+)^2+q_{\perp}^2}\right)^2\theta( Qz_{cut}-\bar n \cdot k)\nn\\
&+&  \frac{g^2 C_F}{8 \pi^2} \left(\frac{\tilde \mu}{m z_{cut}}\right)^{2\epsilon}\frac{1}{\epsilon}
\eea
The two corresponding virtual contributions from diagrams (c) and (d) in Figure~\ref{fig:diagrams}), are scaleless and disappear in dimensional regularization. If we combine all the finite terms we have, 
\begin{equation}
  B_{+}^{(\perp)} = \delta(q^+)\delta^{(2)}(\vec{q}_{\perp}) + \frac{\alpha_s C_F}{\pi} \lbc \frac{1}{\pi}\frac{q_{\perp}^2}{q^+} \frac{\theta( q^+- Qz_{cut})}{ ((q^+)^2m^2/Q^2+q_{\perp}^2)^2} 
  -\ln^2 \lp \frac{\mu}{m z_{cut}}\rp+\ln\lp \frac{\mu}{m z_{cut}}\rp \rbc
\end{equation}

which tells us the natural scale for this function $\mu_B = \mu z_{cut}$.
The corresponding anomalous dimension as defined in Eq.(\ref{eq:unmeasRGE}) is,
\begin{equation}
  \gamma_{B}^{(\perp)} = \frac{\alpha_s C_F}{\pi} \lbc -2 \ln \lp\frac{\mu}{m z_{cut}} \rp + 1 \rbc
\end{equation}


\subsubsection{Region 2: $q_{\perp} \gg m(1-z)$}
\label{BHQETII}
In this case, the HQET mode does not contribute to the measurement of the transverse momentum but otherwise has an identical definition to the previous case.
\bea
B_+ = \frac{1}{\mathcal{N}} \langle 0 | \bar h_{v_+} W_n \delta(q^+-(1-\hat \Theta_{SD})\mathcal{P}^-) W_n^{\dagger} h_{v_+} |0 \rangle . 
\eea
The real and virtual diagrams remain the same as before. The result for diagram (a) is
\bea
B_{+,\;\text{a}} &=&  2 g^2 C_F \tilde \mu^{2\epsilon} \int \frac{d^dk}{(2\pi)^{d-1}}\frac{\delta^+(k^2)\delta(\bar n \cdot k- q^+) \bar n \cdot v_+}{v_+ \cdot k \bar n \cdot k}\theta( \bar n \cdot k - Qz_{cut})\nn\\
&+&  2 g^2 C_F \tilde \mu^{2\epsilon} \delta(q^+)\int \frac{d^dk}{(2\pi)^{d-1}}\frac{\delta^+(k^2) \bar n \cdot v_+}{v_+ \cdot k \bar n \cdot k}\theta(   Qz_{cut}-\bar n \cdot k)\nn\\
&=&  2 g^2 C_F \left(\frac{\tilde \mu}{(m/Q)q^+}\right)^{2\epsilon}\frac{\pi}{(2\pi)^3}\frac{\theta(q^+-Qz_{cut})}{q^+}\Gamma[\epsilon]\nn\\
&-&\frac{g^2 C_F}{8 \pi^2} \delta(q^+)\left(\frac{\tilde \mu}{m z_{cut}}\right)^{2\epsilon}\frac{\Gamma[\epsilon]}{\epsilon}
\eea

and for diagram (b)
\bea
B_{+,\;\text{b}} &=&  -g^2 C_F \int d^dk \frac{\delta(k^2) (v_+)^2 \delta( q^+-\bar n \cdot k)\theta( \bar n \cdot k - Qz_{cut})}{(v_+ \cdot k )^2}\nn\\
&-&g^2 C_F \delta( q^+)\int d^dk \frac{\delta(k^2) (v_+)^2 }{(v_+ \cdot k )^2}\theta(   Qz_{cut}-\bar n \cdot k)\nn\\
&=& -\frac{g^2 C_F}{(2\pi)^3} \frac{\pi\theta( q^+ - Qz_{cut})}{q^+}+ \frac{g^2 C_F}{8 \pi^2} \left(\frac{\tilde \mu}{m z_{cut}}\right)^{2\epsilon}\frac{1}{\epsilon}
\eea
As before the virtual contributions are scale-less and thus vanish in pure dimensional regularization. The sum of all finite terms from the real diagrams is 
\begin{equation}
  \label{eq:Bplusfinite}
  B_+= \delta(q_+)+ \frac{\alpha_s C_F}{\pi} \lbc  \lb 2\ln \lp \frac{\mu Q}{m q^+}  \rp-1 \rb \frac{\theta( q^+ - Qz_{cut})}{q^+} + \lb- \ln^2 \lp  \frac{\mu}{mz_{cut}}\rp + \ln \lp  \frac{\mu}{mz_{cut}}\rp  \rb \delta(q^+) \rbc
\end{equation}
The corresponding anomalous dimension as defined in Eq.(\ref{eq:measRGE}) is
\begin{equation}
  \gamma_{B} = \frac{\alpha_s C_F}{\pi} \lbc -2 \ln \lp\frac{\mu}{m z_{cut}} \rp + 1 +2 \frac{\theta( q^+ - Qz_{cut})}{q^+}\rbc
\end{equation}
which dictates the natural scale of this function to be again $\mu_B =z_{cut}$.
\subsection{HQETm}
\label{HQETm}
We can borrow the HQET matching coefficient from literature \cite{Hoang:2015vua}. The result is known to two loops but we only require the one loop calculation in our paper.
\bea
H_+(m) =  \frac{\alpha_s C_F}{\pi}\left(\ln^2 \left (\frac{\mu}{m}\right)+\frac{1}{2}\ln \left(\frac{\mu}{m}\right)\right)
\eea
which gives AD
\begin{equation}
  \label{eq:gammaplus}
  \gamma_{+} =   2\frac{\alpha_s C_F}{\pi}\ln \left(\frac{\mu}{m}\right)+ \frac{\alpha_s C_F}{2\pi}
\end{equation}
This function obviously lives at the scale $\mu_m= m$.

\subsection{CSoft}
\label{CSoft}
The collinear-soft function is not affected by the mass of the heavy quark. Hence it is the same as in a massless jet. The collinear-soft function is defined by the matrix element 
\bea
S_C(z_{cut},q_{\perp}) = \frac{1}{N_C} Tr \langle 0 |T\left(U_n^{\dagger}W_t\right)\mathcal{M}^{SD}(q^+,\vec{q}_{\perp}) \bar T \left(W_t^{\dagger} U_n\right)|0 \rangle\nn\\
\eea
The collinear-soft modes only contribute to the measurement if they pass soft-drop, which is implemented by the $\hat \Theta_{SD}$ term.
Then the one-loop colinear-soft function is,
\bea
S_{C,\;\text{a}} &=&4 g^2 C_F \tilde \mu^{2\epsilon} \int \frac{d^dk}{(2\pi)^{d-1}}\frac{\delta^+(k^2)\delta(\bar n \cdot k- q^+) \delta^2(\vec{q}_{\perp}-\vec{k}_{\perp})}{n \cdot k \bar n \cdot k}\theta( \bar n \cdot k - Qz_{cut})\nn\\
& +& 4 g^2 C_F \tilde \mu^{2\epsilon}\delta(q^+) \delta^2(\vec{q}_{\perp}) \int \frac{d^dk}{(2\pi)^{d-1}}\frac{\delta^+(k^2)}{n \cdot k \bar n \cdot k}\theta( Qz_{cut}-\bar n \cdot k  ) \nn \\
&\equiv& S_{C,a-1}+ S_{C,a-2}
\eea
$S_{C,a-2}$ is scaleless and disappears in dimensional regularization.
\bea
S_{C,\;\text{a-1}} =  2 g^2 C_F \frac{1}{q_{\perp}^2}\left(\frac{\tilde \mu}{q_{\perp}}\right)^{2\epsilon}\frac{1}{(2\pi)^3}\frac{\theta(q^+-Qz_{cut})}{q^+}
\eea
Expanding in $\epsilon$ and keeping only the finite contributions we get the renormalized collinear-soft function in $\overline{\text{MS}}$
\begin{equation}
  \label{eq:softfinite}
  S_C = \delta(q^+) \delta^{(2)}(\vec{q}_{\perp})+ \frac{2\alpha_s C_F}{\pi} \mathcal{L}_0(q_{\perp},\mu) \frac{\theta(q^+-Qz_{cut})}{q^+}
\end{equation}
where $\mathcal{L}_0$ is a distribution function defined as 
\bea
\mathcal{L}_0(q_{\perp},\mu) = \frac{1}{2\pi}\frac{1}{\mu^2} \left[\frac{\mu^2}{q^2_{\perp} }\right]_+
\eea
The properties of these functions are collected in \cite{Chiu:2012ir}. For our purposes, we need the relation
\bea 
\mu \frac{d}{d \mu} \mathcal{L}_0(q_{\perp},\mu) = - \delta^2(\vec q_{\perp})
\eea
and the corresponding anomalous dimension defined by Eq.(\ref{eq:measRGE}) (in impact parameter space) is
\begin{equation}
  \gamma_{cs} =  -2 \frac{ \alpha_s C_F}{\pi} \frac{\theta(q^+-Qz_{cut})}{q^+}
\end{equation}

\section{Outline of factorization}
\label{sec:factorization}
In this appendix we briefly discuss the factorization formulae in the presence of soft-drop  grooming. We first discuss the generic result of Eq.(\ref{eq:masterfactorization}) and later we focus on the subsequent  refactorizations of the jet function in Eqs.(\ref{eq:R1factorization}) and (\ref{eq:R2factorization}) for regions 1 and 2 respectively. Here we only give the outline of the relevant proof for factorization. For the reader interested in a more rigorous approach we suggest reading first Section 3 of Ref.~\cite{Ellis:2010rwa} and Section 4 of Ref.~\cite{Bauer:2011uc}. We start from the full theory cross section,
\begin{equation}
  \frac{d\sigma}{dz d^2 \vec{q}_{\perp}} =\frac{ 1}{\mathcal{N}} \int dx d^2p_{h \perp}\Big{\vert}_{p_h^2=m^2} \sum_X \Big{\vert} \langle 0 \vert j^{\mu} \mathcal{M}^{SD} (q^+, \vec{q}_{\perp}) \vert X h(x,\vec{p}_{h\perp}) \rangle L_{\mu} \Big{\vert}^2 \delta^{(4)}(Q^{\mu} - p^{\mu}_{Xh}) \;.
\end{equation}
$j ^{\mu}$ is the electromagnetic current for quarks, $ \mathcal{M}^{SD}$ implements the Soft-drop groomed measurement. $L^{\mu}$ is the wavefunction term for $e^+ e^-$ initial state.
Since we are interested in the generic kinematic region $Q\gg Q(1-z) \sim Q z_{cut} \gg \vert \vec{q}_{\perp} \vert $ only soft and collinear modes will contribute to the final state radiation. We thus match to SCET which is the appropriate EFT in this region of phase space. The matching process consists of two major steps:
\begin{itemize}
\item Match the full theory currents $j^{\mu}$ onto the SCET current operators,
  \begin{equation}
    j^{\mu}(x) = \sum_i C_{i}^{\mu} \cdot  \bar{\chi}_{\bar{n}} \Gamma_{i} \chi_n (x)
  \end{equation}
  where $C_i$ are the Wilson short distance matching coefficients.
\item factorize the Hilbert final states onto the subspace of relevant kinematics,
  \begin{equation}
    \vert X+ h(z,\vec{p}_{h \perp}) \rangle \to  \vert X_{reco.}\rangle |X_{coll.}+h(z,\vec{p}_{h \perp}) \rangle |X_{soft} \rangle
  \end{equation}
  where $X_{recol.}$ is the radiation recoiling to the hemisphere jet of interest while $h$ is the final state meson that is identified in the jet. Also since we are considering only the threshold region the heavy meson is also collinear and hence inhabits the same Hilbert space as $X_{coll.}$.
\end{itemize}

It is important to note that soft radiation has transverse momentum parametrically larger than the heavy meson and thus, from the kinetic constraints, doomed to fail the soft-drop grooming. Also important, the recoil radiation, which is collinear along the opposite hemisphere axis, is decoupled from collinear modes in the SCET Lagrangian. In addition, the soft modes can be decoupled from collinear radiation through the BPS field redefinition,
\begin{align}
  \chi_n(x) & = Y_n^{\dag} (x) \chi^{(0)}_n(x)\;, \nn \\
   \bar{\chi}_n(x) & =\bar{ \chi}^{(0)}_n(x)  Y_n(x) \;,
\end{align}
where $\chi^{(0)}_n$ is the uncoupled quark field. Pushing all of he normalization constants, leptonic tensors, Wilson matching coefficients, and recoiling radiation matrix elements into one factor, $\sigma_0(E_J)$ we can write the cross section as follows:
\begin{multline}
  \frac{d\sigma}{dz d^2 \vec{q}_{\perp}} = \sigma_{0}(E_J) \times \frac{1}{N_C} \text{Tr} \langle 0|T\{Y_nY_{\bar n}\}\hat \Theta_{SD}\Theta_R\bar T\{Y_n Y_{\bar n}\}|0\rangle \\
  \times \int dx d^2p_{h \perp}\Big{\vert}_{p_h^2=m^2} \sum_{X_{coll.}} \Big{\vert} \langle 0 \vert \chi_{n} \mathcal{M}^{SD} (q^+, \vec{q}_{\perp}) \vert X_{coll} h(x,\vec{p}_{h\perp}) \rangle  \Big{\vert}^2 \delta^{(4)}(p_{jet}^{\mu} - p^{\mu}_{X_{coll.} h}) \;.
\end{multline}
defining then the second term in the first line as the global soft function and the second line as the jet function we retrieve Eq.(\ref{eq:masterfactorization}).

For further refactorization of the jet function we need to match the collinear  fields of SCET onto SCET$_M$ and subsequently onto bHQET. This processes is discussed in detail in Ref.~\cite{Fleming:2007qr} and is effectively achieved through the following replacement,
\begin{equation}
  \chi_{n} \to C_{+}  W^{\dag}_{n} h_{v_+}
  \end{equation}
	Here $C_+$ is the amplitude level matching cofficient from SCET$_M$ to bHQET. $W^{\dag}_n$ is the Wilson line obtained by decoupling the ultra-collinear mode (which describes the IR fluctuations around a boosted heavy quark) of bHQET from all the other sectors.
Defining $H_+ =C_+ C_+^{*}$ we have the refactorization of the jet function to the hard function $H_+$ and the bHQET jet function as described in Eq.(\ref{eq:R1factorization})

If a hierarchy exist be the scales $q_{\perp}$ and $m(1-z)$ then we are in region 2 and a second collinear mode (referred to as the collinear-soft mode) is relevant to measurement which need to be included in the effective theory. The collinear-soft mode has the following scaling,
\begin{equation}
  p_{sc}^{\mu} \sim (Q(1-z),q_{\perp}^2/Q(1-z),q_{\perp})
\end{equation}
The matching from SCET onto the new effective theory may achieved with the following replacement,
\begin{equation}
  \chi_{n} \to C_{+}  W^{\dag}_{t} W^{\dag}_{n} h_{v_+}
\end{equation}
where the collinear Wilson line $W_{t}$ consist of collinear-soft fields and the matching coefficients $C_{+}$ are the same as in matching onto bHQET.  At this stage the  heavy quark field and the collinear-soft modes are still coupled at the level of the Lagrangian. One needs to perform a BPS-like field redefinition to achieve further refactorization. This is similar to what is done in SCET$_+$ in Ref~\cite{Bauer:2011uc}, i.e.,
\begin{equation}
   W^{\dag}_{n} h_{v_+} = U_n W^{(0)\dag}_{n} h^{(0)}_{v_+}
\end{equation}
where $ W^{(0)\dag}_{n}$ and $ h^{(0)}_{v_+}$ are decoupled from the collinear-soft fields.
To proceed we need to apply power-counting to the measurement operator $\mathcal{M}^{SD}$ and more particularly to the operators $\mathcal{P}^-$ and $\mathcal{P}_{\perp}$. While the contribution of ultra-collinear and collinear-soft modes to the energy fraction is similar the contribution of the ultra-collinear in perpendicular component is suppressed,
\begin{align}
  \mathcal{P}^- &\sim \mathcal{P}^-_{cs} +\mathcal{P}^-_{uc}\nn\\
  \mathcal{P}_{\perp} &\sim \mathcal{P}_{\perp cs}
\end{align}
With these modifications is then trivial to show the refactorization of the bHQET jet function into an ultra-collinear component and the collinear soft function as shown in Eq.(\ref{eq:R2factorization}).

\bibliography{./bjet}
\end{document}